\documentclass[12pt]{article}
\usepackage{amsfonts}
\usepackage{amssymb}
\usepackage{amsbsy}
\usepackage{graphics}
\usepackage{epsfig}

\input epsf.sty

\newlength{\TZ}
\newcommand{\BEQ}{\begin{equation}}     
\newcommand{\BEA}{\begin{eqnarray}}
\newcommand{\EEQ}{\end{equation}}       
\newcommand{\EEA}{\end{eqnarray}}
\newcommand{\eps}{\varepsilon}          
\newcommand{\D}{{\rm d}}                
\newcommand{\II}{{\rm i}}               
\newcommand{\demi}{\frac{1}{2}}         
\newcommand{\wit}[1]{\widetilde{#1}}    
\newcommand{\rar}{\rightarrow}          

\renewcommand{\vec}[1]{\boldsymbol{#1}} 

\newcommand{\vekz}[2]
     {\mbox{${\begin{array}{c} #1  \\ #2 \end{array}}$}}
\newcommand{\matz}[4] 
     {\mbox{${\begin{array}{cc} #1 & #2 \\ #3 & #4 \end{array}}$}}

\newcommand{\appsection}[2]{\setcounter{equation}{0}\setcounter{subsection}{0}
\section*{Appendix #1. #2}
\renewcommand{\theequation}{#1.\arabic{equation}}
              \renewcommand{\thesection}{#1} }

\catcode`\@=11
\def\numberbysection{\@addtoreset{equation}{section}
        \def\theequation{\thesection.\arabic{equation}}}
\numberbysection

\begin{document}

\begin{titlepage}

\vskip 1.5 cm
\begin{center}
{\LARGE \bf On logarithmic extensions of local scale-invariance}
\end{center}

\vskip 2.0 cm
\centerline{{\bf Malte Henkel}}
\vskip 0.5 cm
\centerline{Groupe de Physique Statistique,
D\'epartement de Physique de la Mati\`ere et des Mat\'eriaux,}
\centerline{Institut Jean Lamour (CNRS UMR 7198), 
Universit\'e de Lorraine Nancy,} 
\centerline{B.P. 70239, F -- 54506 Vand{\oe}uvre l\`es Nancy Cedex, France}

\begin{abstract}
Ageing phenomena far from equilibrium naturally present dynamical scaling and in many situations
this may generalised to local scale-invariance. 
Generically, the absence of time-translation-invariance implies
that each scaling operator is characterised by two independent scaling dimensions. 
Building on analogies with logarithmic conformal invariance and logarithmic Schr\"odinger-invariance, 
this work proposes a logarithmic extension of local scale-invariance, 
without time-translation-invariance. 
Carrying this out requires in general to replace both scaling dimensions 
of each scaling operator by Jordan cells. 
Co-variant two-point functions are derived for the 
most simple case of a two-dimensional logarithmic extension. 
Their form is compared to simulational data for autoresponse functions 
in several universality classes of non-equilibrium ageing phenomena.  
\end{abstract}

\end{titlepage}

\setcounter{footnote}{0} 

\section{Introduction}
Scale-invariance has become one of the main characteristics of phase transitions and
critical phenomena. In many situations, especially when in the case of sufficiently local 
interactions, scale-invariance can be extended to larger Lie groups of coordinate transformations. 
For the analysis of phase transitions at equilibrium, conformal invariance has played a central r\^ole, 
especially in two spatial dimensions \cite{Polyakov70,Belavin84}. It is then natural to inquire into the equilibrium
critical dynamics at a critical point, where the spatial dilatations $\vec{r}\mapsto \lambda \vec{r}$ 
are extended to include a temporal
dilatation as well, viz. $t\mapsto \lambda^z t$, $\vec{r}\mapsto \lambda \vec{r}$, and where the
dynamical exponent $z$ describes the distinct behaviour of time with respect to space. Indeed, it was attempted
to use $2D$ conformal invariance in  this context \cite{Cardy85}. 

However, known results concerning the dynamical symmetries of free diffusion (or Schr\"odinger) equations suggested 
a different line of inquiry. It has been known since the 18$^{\rm th}$ 
century to mathematicians as Lie and Jacobi \cite{Lie1881,Jacobi1843} 
that the following set of space-time transformations
\BEQ \label{sch1}
t \mapsto \frac{\alpha t +\beta}{\gamma t+\delta} \;\; , \;\; 
\vec{r} \mapsto \frac{{\cal R}\vec{r} + \vec{v} t + \vec{a}}{\gamma t+\delta} \;\; ; \;\;
\alpha\delta - \beta\gamma=1
\EEQ
maps any solution of the free diffusion or Schr\"odinger equation onto another solution of the same equation,
provided the wave function is transformed accordingly with a known projective factor. This makes up the so-called
{\em Schr\"odinger group} {\sl Sch}(d), and with its Lie algebra denoted by $\mathfrak{sch}(d)$. Herein, the transformations
are parametrised by ${\cal R}\in{\sl SO}(d)$, $\vec{a},\vec{v}\in\mathbb{R}^d$ and $\alpha,\beta,\gamma,\delta\in\mathbb{R}$. 
Clearly, the Schr\"odinger group includes conformal transformations in the {\em time} $t$, and the spatial transformations admitted
are selected in order to close the group product (or the Lie algebra commutators). In particular, one may read off the
dynamical exponent $z=2$, by considering in the representation (\ref{sch1}) the
dilatations (where $\beta=\gamma=0$, $\vec{v}=\vec{a}=\vec{0}$ and ${\cal R}={\bf 1}$). Therefore, the Schr\"odinger group
might be seen as an useful starting point for studying consequences of dynamical scaling, where $z\ne 1$. 

During the last decade, the relevance of the Schr\"odinger algebra to {\em non}-equilibrium dynamical scaling has become 
increasingly clear. In contrast to equilibrium critical scaling, which requires the fine-tuning of physical parameters to
a well-defined critical point, {\em dynamical scaling} 
may arise naturally in a large variety of many-body systems far from equilibrium,
and without having to fine-tune physical parameters. 

A paradigmatic example of non-equilibrium dynamics are {\em ageing phenomena}. 
An often-studied realisation of ageing may arise in systems which are initially
prepared in a high-temperature initial state, by bringing them into contact with a heat-bath. 
The system is then brought out of equilibrium by rapidly  changing the heat-bath temperature rapidly to low values (`quenching'), 
either (a) into a coexistence phase with more than one stable equilibrium state or else 
(b) onto a critical point of the stationary state \cite{Bray94a,Cugliandolo02,Henkel10}. Based on many experimental observations
and numerical studies of models, it has emerged that from a phenomenological
point of view, ageing can be defined through the properties:
\begin{enumerate}
\item slow, non-exponential relaxation, 
\item breaking of time-translation-invariance, 
\item dynamical scaling. 
\end{enumerate}
Although ageing was first systematically studied in glassy systems, 
where the dynamics is characterised by the simultaneous effects of
both disorder and frustrations, very similar phenomena have also been found even in quenched simple magnets 
(ferromagnetic, without disorder).\footnote{At this stage, several distinct types of dynamical scaling, 
corresponding to {\em full ageing} (e.g. in simple magnets) or {\em sub-ageing} 
(e.g. in glassy systems), remain possible. In this paper, only models with full ageing are considered.} 

A possible use of dynamical scaling is suggested by drawing an analogy with
equilibrium critical phenomena, where scale-invariance can often be extended
to conformal invariance \cite{Polyakov70,Belavin84} (under rather weak conditions). One of the first applications
of such an approach is the prediction of some elementary two- and three-point functions of the quasi-primary scaling
operators in a given theory. Therefore, one may ask whether analogous extensions of simple dynamical scaling with
a dynamical exponent $z$ might exist. If that is so, such a dynamical symmetry could be called
{\em local scale-invariance ({\sc lsi})}. Applied to ageing, it is clear that the full Schr\"odinger algebra
$\mathfrak{sch}(d)$ {\em cannot} be used, even if $z=2$. Rather, one should consider the sub-algebra obtained when leaving out
the time-translation generator. This algebra will be called {\em ageing algebra} and is denoted by $\mathfrak{age}(d)$. 
Since ageing systems are far from equilibrium, there no longer exists a fluctuation-dissipation theorem which could
relate correlators and responses. It turns out that far from equilibrium the response functions transform covariantly under
$\mathfrak{age}(d)$ -- and in contrast to conformal invariance at equilibrium, ageing-invariance is needed to fix the form
of an universal, but non-trivial scaling function. Indeed, the form of the linear
two-time autoresponse function of the order-parameter $\phi(t,\vec{r})$ 
with respect to its canonically conjugated external field $h(s,\vec{r})$, in the scaling limit where 
$s\to\infty$, $0<t-s\to\infty$ such that $y=t/s$ is kept fixed, reads 
\cite{Henkel01b,Henkel03a,Henkel06a,Henkel12a}
\BEA 
R(t,s) &=& \left.\frac{\delta \langle\phi(t,\vec{r})\rangle}{\delta h(s,\vec{r})}\right|_{h=0} 
\:=\: \left\langle \phi(t,\vec{r}) \wit{\phi}(s,\vec{r}) \right\rangle 
\:=\: s^{-1-a} f_R\left(\frac{t}{s}\right) \;\; , \;\; 
\nonumber \\
f_R(y) &=& f_0 y^{1+a'-\lambda_R/z} (y-1)^{-1-a'} \Theta(y-1)
\label{R}
\EEA
where $\langle .\rangle$ denotes a thermodynamic average. The re-writing of the response $R$ as a correlator
between the order-parameter $\phi$ and an associated {\em `response field'} 
$\wit{\phi}$ is a well-known consequence of Janssen-de Dominicis theory \cite{Janssen92,Cugliandolo02}. 
In eq.~(\ref{R}), the autoresponse exponent $\lambda_R$ and the ageing exponents $a,a'$ are universal
non-equilibrium exponents.\footnote{In simple magnets, mean-field theory suggests that generically
$a=a'$ for quenches to $T<T_c$ and $a\ne a'$ for $T=T_c$ \cite{Henkel10}. 
Hence co-variance under $\mathfrak{age}(d)$ is required for deriving (\ref{R}), 
whereas $\mathfrak{sch}(d)$-covariance would produce therein the extra constraint $a=a'$ \cite{Henkel06a}.}
The causality condition $y=t/s>1$ is explicitly included via a Theta function. 
The foundations and extensive tests of (\ref{R}) are reviewed in detail in \cite{Henkel10}. 

Clearly, a prediction such as (\ref{R}) can merely provide a first step towards a construction of a fully local
form of dynamical scaling. Although eq.~(\ref{R}) is indeed very well reproduced in several exactly solved models, as well
as in many simulational studies, we shall describe in section~4 that in certain models of non-equilibrium ageing, 
the scaling function given in (\ref{R})
only captures partially the model data. In this work, we describe a possible extension of {\sc lsi}, which draws on one side on 
specific features of the representation theory of the ageing algebra $\mathfrak{age}(d)$, 
coming from the absence of time-translation-invariance, and on the other hand is inspired
by the well-known logarithmic extensions of conformal invariance. In the remainder of this introduction, we shall recall this latter
aspect, before we construct logarithmic extensions of $\mathfrak{age}(d)$ and discuss some applications in the later sections. 

In the case of a degenerate vacuum state, 
conformal invariance (of equilibrium phase transitions) can be generalised to 
{\em logarithmic conformal invariance} \cite{Gurarie93,Gaberdiel96,Rahimi97,Gurarie99,Gurarie04}, 
with interesting applications
to disordered systems \cite{Caux96,Maassarani97}, percolation \cite{Flohr05,Mathieu07,Vasseur12b} 
sand-pile models \cite{Poghosyan07}, or critical spin systems \cite{Vasseur12a}. 
For reviews, see \cite{Flohr03,Gaberdiel03}. 
Here, we shall be interested in possible logarithmic extensions of 
local scale-invariance and in the corresponding generalisations of (\ref{R}). 

Logarithmic conformal invariance in $2D$ can be heuristically introduced \cite{Gurarie93,Rahimi97} 
by replacing, in the left-handed chiral conformal
generators $\ell_n = - w^{n+1}\partial_z - (n+1) w^n \Delta $, 
the conformal weight $\Delta$ by a Jordan
matrix.\footnote{Throughout, the complex coordinates $w=w_x+\II w_y$ will be used, 
in order to avoid possible confusion with the dynamical exponent $z$.} 
Non-trivial results are only obtained if that matrix has a Jordan form, so that one writes,
in the most simple case
\BEQ \label{1.1}
\ell_n = - w^{n+1} \partial_w - (n+1) w^n \left(\matz{\Delta}{1}{0}{\Delta}\right)
\EEQ
Then the quasi-primary scaling operators on which the $\ell_n$ act have two components, 
which we shall
denote as $\Psi :=\left(\vekz{\psi}{\phi}\right)$. 
The generators (\ref{1.1}) satisfy the commutation relations
$[\ell_n, \ell_m] = (n-m) \ell_{n+m}$ with $n,m\in\mathbb{Z}$. 
Similarly, the right-handed generators $\bar{\ell}_n$
are obtained by replacing $w\mapsto \bar{w}$ and $\Delta\mapsto \bar{\Delta}$. 
A simple example of an invariant equation can be
written as ${\cal S}\Psi=0$, with the Schr\"odinger operator
\BEQ
{\cal S} := \left( \matz{0}{\partial_w \partial_{\bar{w}}}{0}{0} \right)
\EEQ
Because of 
$[{\cal S},\ell_n] = - (n+1) w^n {\cal S} - 
(n+1)n w^{n+1} \left( \matz{0}{\Delta}{0}{0}\right) \partial_{\bar{w}}$, 
and if one chooses the conformal weights $\Delta=\overline{\Delta}=0$, the generators (\ref{1.1}) 
act as dynamic symmetries in that solutions of
the equation ${\cal S}\Psi=0$ are mapped onto other solutions. 

Of particular importance are the consequences for the form of 
the two-point functions of quasi-primary operators, for which
only co-variance under the finite-dimensional sub-algebra 
$\langle \ell_{\pm 1,0}\rangle \cong \mathfrak{sl}(2,\mathbb{R})$ is
needed \cite{Gurarie93,Rahimi97} 
(we suppress the dependence on $\bar{w}_i$, but see \cite{Do08}). Set
\BEQ \label{1.3}
F := \left\langle \phi_1(w_1) \phi_2(w_2)\right\rangle \;\; , \;\; 
G := \left\langle \phi_1(w_1) \psi_2(w_2)\right\rangle \;\; , \;\; 
H := \left\langle \psi_1(w_1) \psi_2(w_2)\right\rangle
\EEQ 
Translation-invariance implies that 
$F=F(w), G=G(w)$ and $H=H(w)$ with $w=w_1-w_2$. 
Combination of dilation- and special
co-variance applied to $F,G$ leads to $\Delta :=\Delta_1=\Delta_2$ and $F(w)=0$. 
Finally, consideration of $H(w)$ leads to
\BEQ \label{1.4}
G(w) = G(-w) = G_0 |w|^{-2\Delta} \;\; , \;\; 
H(w) = H(-w) = \bigl( H_0 - 2 G_0 \ln |w|\bigr) \, |w|^{-2\Delta}
\EEQ
where $G_0, H_0$ are normalisation constants. 
We emphasise here the {\em symmetric} form of the two-point functions, 
which does follow from the three co-variance conditions (see appendix~A for a reminder). 

Recently, `non-relativistic' versions of logarithmic conformal invariance have 
been studied \cite{Hosseiny10}. 
Besides the consideration of dynamics in statistical physics referred to above, 
such studies can also be motivated from the 
analysis of dynamical symmetries in non-linear hydrodynamical
equations \cite{Ovsiannikov80,Niederer78,Ivash97,Hassaine00,ORaif01}, 
or from studies of non-relativistic versions of the AdS/CFT 
correspondence \cite{Maldacena98,Bala08,Son08,Minic08,Fuertes09,Leigh09,Hartnoll09,Minic12}. 
Two distinct non-semi-simple Lie algebras have been considered:
\begin{enumerate}
\item the {\em Schr\"odinger algebra} $\mathfrak{sch}(d)$, 
identified in 1881 by Lie as maximal dynamical symmetry of the
free diffusion equation in $d=1$ dimensions. 
Jacobi had observed already in the 1840s that the elements of $\mathfrak{sch}(d)$ 
generate dynamical symmetries of free motion. We write the generators compactly as follows
\BEA
X_n &=& -t^{n+1}\partial_t - \frac{n+1}{2}t^n \vec{r}\cdot\vec{\nabla}_{\vec{r}} 
- \frac{\cal M}{2}(n+1)n t^{n-1} \vec{r}^2
- \frac{n+1}{2} x t^n \nonumber \\
Y_m^{(j)} &=& - t^{m+1/2} \partial_j - \bigl( m + \demi\bigr) t^{m-1/2} r_j \nonumber \\
M_n &=& - t^n {\cal M} \label{1.5} \\
R_n^{(jk)} &=& - t^n \bigl( r_j \partial_k - r_k \partial_j \bigr) \nonumber
\EEA
where $\cal M$ is a dimensionful constant, $x$ a scaling dimension, 
$\partial_j = \partial/\partial r_j$ and $j,k=1,\ldots,d$. Then 
$\mathfrak{sch}(d)=\langle X_{\pm 1,0}, Y_{\pm 1/2}^{(j)}, M_0, R_0^{(j,k)}\rangle_{j,k=1,\ldots,d}$ 
is a dynamical symmetry of the free Schr\"odinger equation
${\cal S}\phi = \bigl( 2{\cal M}\partial_t - \vec{\nabla}_{\vec{r}}^2\bigr)\phi=0$, provided $x=d/2$, 
see \cite{Kastrup68,Hagen72,Niederer72,Jackiw72}, 
and also of Euler's hydrodynamical equations \cite{Ovsiannikov80}. 
An infinite-dimensional extension is
$\mathfrak{sv}(d) := \langle X_n, Y_m^{(j)}, M_n, 
R_0^{(jk)}\rangle_{n\in\mathbb{Z},m\in\mathbb{Z}+\demi,j,k=1,\ldots,d}$ 
\cite{Henkel94}, with applications to Burger's equation \cite{Ivash97}. 
\item The Schr\"odinger algebra is {\em not} 
the non-relativistic limit of the conformal algebra. Rather, from the
corresponding contraction one obtains the 
{\em conformal Galilei algebra} $\mbox{\sc cga}(d)$ \cite{Havas78}, 
which was re-discovered independently
several times afterwards \cite{Henkel97,Negro97,Henkel03a,Bagchi09,Martelli09}. 
The generators may be written as follows \cite{Cherniha10}
\BEA X_n &=&
- t^{n+1}\partial_t - (n+1) t^n \vec{r}\cdot\vec{\nabla}_{\vec{r}}
- n(n+1) t^{n-1} \vec{\gamma}\cdot\vec{r} - x (n+1)t^n
\nonumber \\
Y_n^{(j)} &=& - t^{n+1} \partial_{j} - (n+1) t^n \gamma_j  \label{1.6} \\
R_n^{(jk)} &=& - t^n \bigl( r_j \partial_{k} -  r_k \partial_{j} \bigr)
- t^n \bigl( \gamma_j \partial_{\gamma_k}-\gamma_k
\partial_{\gamma_j}\bigr) \nonumber 
\EEA
where $\vec{\gamma}=(\gamma_1,\ldots,\gamma_d)$ 
is a vector of dimensionful constants and $x$ is again a scaling dimension. 
The algebra 
$\mbox{\sc cga}(d)=\langle X_{\pm 1,0},Y_{\pm 1,0}^{(j)},R_0^{(jk)}\rangle_{j,k=1,\ldots,d}$ 
does arise as a 
(conditional) dynamical symmetry in certain non-linear systems, 
distinct from the equations of non-relativistic 
incompressible fluid dynamics \cite{Zhang10,Cherniha10}.\footnote{The generator $X_0$ 
leads to the space-time dilatations $t\mapsto \lambda^z t$, $\vec{r}\mapsto \lambda \vec{r}$, 
where the dynamical exponent $z$ takes the value $z=2$ for the representation (\ref{1.5}) of 
$\mathfrak{sch}(d)$ and $z=1$ for the representation (\ref{1.6}) of $\mbox{\sc cga}(d)$. We point 
out that there exist representations of $\mbox{\sc cga}(d)$ with $z=2$ \cite{Henkel03a}.
{}From this, one can show that $\mathfrak{age}(1)\subset\mbox{\sc cga}(1)$ as well.} 
The infinite-dimensional extension
$\mathfrak{av}(d) := \langle X_{n},Y_{n}^{(j)},R_n^{(jk)}\rangle_{n\in\mathbb{Z},j,k=1,\ldots,d}$ is straightforward. 
\end{enumerate}
For both algebras $\mathfrak{sch}(d)$ and
 $\mbox{\sc cga}(d)$, the non-vanishing commutators are given by
\BEQ \label{1.7}
{}[X_n, X_{n'}] = (n-n') X_{n+n'} \;,\; 
{}[X_n, Y_{m}^{(j)}] = \left(\frac{n}{z}-m\right)Y_{n+m}^{(j)} \;,\;
{}[R_0^{(jk)}, Y_m^{(\ell)}] = \delta^{j,\ell} Y_m^{(k)} - \delta^{k,\ell} Y_m^{(j)}
\EEQ
together with the usual commutators of the rotation group $\mathfrak{so}(d)$, and 
where the dynamical exponent $z=2$ for the representation (\ref{1.5}) of $\mathfrak{sch}(d)$ and 
$z=1$ for the representation (\ref{1.6}) of $\mbox{\sc cga}(d)$. 
For the Schr\"odinger algebra $\mathfrak{sch}(d)$, 
one has in addition $[Y_{1/2}^{(j)}, Y_{-1/2}^{(k)}] = \delta^{j,k} M_0$. 
  
The algebras $\mathfrak{sch}(d)$ and $\mbox{\sc cga}(d)$ arise, 
besides the conformal algebra, as the only possible finite-dimensional Lie algebras 
in two classification schemes of non-relativistic space-time transformations, 
with a fixed dynamical exponent $z$, 
namely: (i) either as generalised conformal transformations \cite{Duval09} or
(ii) as local scale-transformations which are conformal in time \cite{Henkel02}. 
 
Now, using the same heuristic device as for logarithmic conformal invariance and 
replacing in the generators $X_n$ in
(\ref{1.5},\ref{1.6}) the scaling dimension by a Jordan matrix
\BEQ
x \mapsto \left(\matz{x}{1}{0}{x}\right)
\EEQ
both {\em logarithmic Schr\"odinger-invariance} and 
{\em logarithmic conformal galilean invariance} can be defined 
\cite{Hosseiny10}. Adapting the definition (\ref{1.3}), 
one now has $F=F(t,\vec{r})$, $G=G(t,\vec{r})$ and $H=H(t,\vec{r})$, with
$t:=t_1-t_2$ and $\vec{r}:=\vec{r}_1-\vec{r}_2$ because of temporal and 
spatial translation-invariance.  
Since the conformal properties involve the time coordinate only, 
the practical calculation is analogous to the one
of logarithmic conformal invariance outlined in appendix~A 
(alternatively, one may use the formalism of nilpotent variables 
\cite{Moghimi00,Hosseiny10}). In particular, one obtains $x:= x_1 = x_2$ and $F=0$.
Generalising the results of Hosseiny and Rouhani \cite{Hosseiny10} to $d$ spatial dimensions, 
the non-vanishing two-point functions read as follows: 
for the case of logarithmic Schr\"odinger invariance 
\BEQ \label{1.8}
G = G_0 |t|^{-x}\exp\left[-\frac{{\cal M}}{2} \frac{\vec{r}^2}{t}\right] \;\; ,\;\;  
H = \bigl( H_0 -  G_0 \ln |t|\bigr) \, |t|^{-x} 
\exp\left[-\frac{{\cal M}}{2} \frac{\vec{r}^2}{t}\right] 
\EEQ
subject to the constraint \cite{Bargman54} 
${\cal M}:= {\cal M}_1 = - {\cal M}_2$.\footnote{In order to keep the
physical convention of non-negative masses ${\cal M}\geq 0$, 
one may introduce a `complex conjugate' $\phi^*$ to the
scaling field $\phi$, with ${\cal M}^*=-{\cal M}$. 
In dynamics, co-variant two-point functions are interpreted 
as response functions, written as $R(t,s)=\left\langle \phi(t) \wit{\phi}(s)\right\rangle$ 
in the context of Janssen-de Dominicis theory, 
where the response field $\wit{\phi}$ has a mass $\wit{\cal M}=-{\cal M}$,
see e.g. \cite{Cugliandolo02,Henkel10} for details.\\
Furthermore, the physical relevant equations are {\em stochastic} 
Langevin equations, whose noise terms do break any interesting extended 
dynamical scale-invariance. However, one may identify a 
`deterministic part' which may be Schr\"odinger-invariant, such that the predictions
(\ref{1.8}) remain valid even in the presence of noise \cite{Picone04}. 
This was rediscovered recently under name of
`time-dependent deformation of Schr\"odinger geometry' \cite{Nakayama10}.} 
For the  case of logarithmic conformal galilean invariance, we have in an analogous way 
\BEQ \label{1.9}
G = G_0 |t|^{-2x}\exp\left[-2\frac{\vec{\gamma}\cdot\vec{r}}{t}\right] \;\;,\;\;  
H = \bigl( H_0 - 2 G_0 \ln |t|\bigr)\, |t|^{-2x} 
\exp\left[-2\frac{\vec{\gamma}\cdot\vec{r}}{t}\right] 
\EEQ
together with the constraint $\vec{\gamma} :=\vec{\gamma}_1 = \vec{\gamma}_2$. Here, 
$G_0,H_0$ are again normalisation constants.\footnote{There is a 
so-called `exotic' central extension of $\mbox{\sc cga}(2)$ \cite{Lukierski06},
 but the extension of the known two-point functions \cite{Bagchi09b,Bagchi09c,Martelli09} to
the logarithmic version has not yet been attempted.}  
The causality condition $t>0$ can be derived, 
for both (\ref{1.8}) and (\ref{1.9}), after a dualisation of the
`mass parameters' quite analogous to the AdS/CFT correspondence, by extending the postulated
symmetry to a maximal parabolic sub-algebra of the (complex) conformal algebra 
$\mathfrak{conf}(d+2)$ in $d+2$
dimensions, see \cite{Henkel12a} for the detailed proof. 
Because of this causality, the most natural physical interpretation
of co-variant two-point functions is in terms of {\em responses}, rather than correlators. 
We shall adopt this point of view in section 4 below. 

{}From the comparison of the results (\ref{1.8},\ref{1.9}) 
with the form (\ref{1.5}) of logarithmic conformal invariance,
we see that logarithmic corrections to scaling are systematically present. 
As we shall show, this feature is a 
consequence of the assumption of time-translation-invariance, since
the time-translation operator $X_{-1}=-\partial_t$ is contained in both algebras. 
On the other hand, from the point of view of non-equilibrium statistical physics, neither the
Schr\"odinger nor the conformal Galilei algebra is a 
satisfactory choice for a dynamical symmetry, since time-translation-
invariance can only hold true at a 
stationary state and hence eqs.~(\ref{1.5},\ref{1.6}) can only be valid 
in situations such as {\em equilibrium} critical dynamics. For non-equilibrium systems, 
it is more natural to leave out time-translations from the algebra altogether. An enormous
variety of physical situations with a natural dynamical scaling is known to exist, 
although the associated stationary state(s), towards which the system is relaxing to, 
need not be scale-invariant \cite{Henkel10}. 
We then arrive at the so-called {\em ageing algebra} 
$\mathfrak{age}(d) := 
\langle X_{0,1},Y_{\pm 1/2}^{(j)}, M_0, R_0^{(jk)}\rangle_{j,k=1,\ldots,d}\subset \mathfrak{sch}(d)$ 
and shall study the consequences of a logarithmic extension of ageing-invariance, 
to which we shall also refer as {\em logarithmic {\sc lsi}} or {\sc llsi} for short.  

In section~2, the generators of logarithmic 
ageing-invariance will be specified and we shall see that an essential distinction from logarithmic conformal or Schr\"odinger invariance 
is that each scaling operator is characterised by {\em two} independent scaling dimensions, 
which will have to be replaced by a Jordan matrix. The co-variant
two-point functions will be derived in section~3. 
In section~4, some possible applications to ageing phenomena will be discussed. 
We shall see that the scaling of the two-time autoresponse function 
in non-equilibrium ageing phenomena can be well fitted to
the predictions of {\sc llsi}. We conclude in section~5. 
Appendix~A recalls the derivation of
two-point function in logarithmic conformal invariance and 
appendix~B shows that a logarithmic scaling form frequently encountered
in ageing phenomena is distinct from logarithmic {\sc lsi}. 

\section{Logarithmic extension of the ageing algebra $\mathfrak{age}(d)$}

For definiteness, consider the {\em ageing algebra} 
$\mathfrak{age}(d) = 
\langle X_{0,1},Y_{\pm 1/2}^{(j)}, M_0, R_0^{(jk)}\rangle_{j,k=1,\ldots,d}\subset \mathfrak{sch}(d)$, 
which is a sub-algebra of the Schr\"odinger algebra. 
The generators of the representation (\ref{1.5}) can in general be taken over, 
but with the important exception 
\BEQ \label{2.1}
X_n = -t^{n+1}\partial_t - \frac{n+1}{2}t^n \vec{r}\cdot\vec{\nabla}_{\vec{r}} 
- \frac{\cal M}{2}(n+1)n t^{n-1} \vec{r}^2
- \frac{n+1}{2} x t^n - (n+1)n \xi t^n 
\EEQ
where now $n\geq 0$ and (\ref{1.7}) remains valid. 
In contrast to the representation (\ref{1.5}), 
one now has {\em two distinct} scaling dimensions $x$ and 
$\xi$, with important consequences on the form of 
the co-variant two-point functions \cite{Picone04,Henkel06a}, 
to be derived below.\footnote{If one assumes time-translation-invariance, the commutator 
$[X_1,X_{-1}]=2X_0$ leads to $\xi=0$ and one
is back to (\ref{1.5}). Physical examples with $\xi\ne 0$ are mentioned below.} To simplify the discussion, 
we shall concentrate from now on the temporal part 
$\langle \Psi(t_1,\vec{r}_0)\Psi(t_2,\vec{r}_0)\rangle$, 
the form of which is described by the two generators $X_{0,1}$, with
the commutator $[X_1,X_0]=X_1$. At the end, the spatial part is easily added. 

Logarithmic representation of $\mathfrak{age}(d)$, 
analogously to section~1, can be constructed by considering two
scaling operators, with {\em both} scaling dimensions $x$ and $\xi$ identical, and replacing
\BEQ \label{2.2}
x \mapsto \left(\matz{x}{x'}{0}{x}\right) \;\; , \;\;
\xi  \mapsto \left(\matz{\xi}{\xi'}{\xi''}{\xi}\right)
\EEQ
in eq.~(\ref{2.1}). The other generators (\ref{1.5}) are kept unchanged. 
Without restriction of generality, one can always achieve either a diagonal form (with $x'=0$) 
or a Jordan form (with $x'=1$) of the first matrix, 
but for the moment it is not yet clear if the second matrix in 
(\ref{2.2}) will have any particular structure. Setting
$\vec{r}=\vec{0}$, we have from (\ref{2.1}) the two generators
\BEQ \label{2.3}
X_0 = - t\partial_t - \demi \left(\matz{x}{x'}{0}{x}\right) \;\; , \;\;
X_1 = - t^2\partial_t - t \left(\matz{x+\xi}{x'+\xi'}{\xi''}{x+\xi}\right) 
\EEQ
and we find $[X_1,X_0]=X_1 +\demi t \,x'\xi'' \left(\matz{-1}{0}{0}{1}\right) \stackrel{!}{=} X_1$. 
The condition $x' \xi''\stackrel{!}{=}0$ follows and we must distinguish two cases.
\begin{enumerate}
\item $x'=0$. The first matrix in (\ref{2.2}) is diagonal. 
In this situation, there are two distinct possibilities: (i) either, the matrix
$\left(\matz{\xi}{\xi'}{\xi''}{\xi}\right)\rar\left(\matz{\xi_+}{0}{0}{\xi_-}\right)$ 
is diagonalisable. One then has
a pair of quasi-primary operators, with scaling dimensions $(x,\xi_+)$ and $(x,\xi_-)$. 
This reduces to the standard
form of non-logarithmic local scale-invariance \cite{Henkel06a}.  Or else, (ii), the matrix 
$\left(\matz{\xi}{\xi'}{\xi''}{\xi}\right)\rar\left(\matz{\bar{\xi}}{1}{0}{\bar{\xi}}\right)$ 
reduces  to a Jordan form.
This is a special case of the situation considered below.
\item $\xi''=0$. Both matrices in (\ref{2.2}) 
reduce simultaneously to a Jordan form. While one can always normalise such 
that either $x'=1$ or else $x'=0$, there is no obvious normalisation for $\xi'$. 
This is the main case which we shall study in  the remainder of this paper.
\end{enumerate}
In conclusion: {\em without restriction on the generality, 
one can set $\xi''=0$ in eqs.~(\ref{2.2},\ref{2.3}).} 

For illustration and completeness, we give an example of a logarithmically 
invariant Schr\"odinger equation. Consider the
Schr\"odinger operator
\BEQ
{\cal S} := \left( 2{\cal M}\partial_t - \vec{\nabla}_{\vec{r}}^2 +\frac{2{\cal M}}{t}
\left( x + \xi - \frac{d}{2}\right) \right) \left( \matz{0}{1}{0}{0}\right)
\EEQ
Using (\ref{2.3}) with the spatial parts restored, 
we have $[{\cal S},X_0]=-{\cal S}$ and $[{\cal S},X_1] = -2t {\cal S}$ and furthermore, 
$\cal S$ commutes with
all other generators of $\mathfrak{age}(d)$. 
Therefore, the elements of $\mathfrak{age}(d)$ map any solution
of ${\cal S}\left(\vekz{\psi}{\phi}\right)=\left(\vekz{0}{0}\right)$ 
to another solution of the same equation. 

\section{Two-point functions}

Consider the following two-point functions, built from the components of quasi-primary operators
of logarithmic {\sc lsi}
\BEA
F = F(t_1, t_2) &:=& \left\langle \phi_1(t_1)\phi_2(t_2)\right\rangle \nonumber \\
G_{12} = G_{12}(t_1, t_2) &:=& \left\langle \phi_1(t_1)\psi_2(t_2)\right\rangle \nonumber \\
G_{21} = G_{21}(t_1, t_2) &:=& \left\langle \psi_1(t_1)\phi_2(t_2)\right\rangle  \\
H = H(t_1, t_2) &:=& \left\langle \psi_1(t_1)\psi_2(t_2)\right\rangle \nonumber 
\EEA
Their co-variance under the representation (\ref{2.3}), with $\xi''=0$, is expressed by the conditions 
$\hat{X}^{[2]}_{0,1}F\stackrel{!}{=}0$,\ldots,
where $\hat{X}^{[2]}_{0,1}$ stands for the extension of (\ref{2.3}) to two-body operators. 
This leads to the following system of eight equations for a set of four functions in two variables. 
\newpage \typeout{*** Seitenvorschub hier ***} 
\BEA
\left[ t_1 \partial_1 + t_2\partial_2 +\demi(x_1+x_2) \right] F(t_1,t_2) &=& 0 
\nonumber \\
\Bigl[ t_1^2 \partial_1 + t_2^2\partial_2 + (x_1+\xi_1) t_1 + (x_2+\xi_2) t_2 \Bigr] F(t_1,t_2) &=& 0 
\nonumber \\[0.20cm]
\left[ t_1 \partial_1 + t_2\partial_2 +\demi(x_1+x_2) \right] G_{12}(t_1,t_2) 
+\frac{x_2'}{2} F(t_1,t_2) &=& 0 
\nonumber \\
\Bigl[ t_1^2 \partial_1 + t_2^2\partial_2 + (x_1+\xi_1) t_1 + (x_2+\xi_2) t_2 \Bigr] G_{12}(t_1,t_2) 
+ (x_2'+\xi_2') t_2 F(t_1,t_2) &=& 0 
\nonumber \\[0.20cm]
\left[ t_1 \partial_1 + t_2\partial_2 +\demi(x_1+x_2) \right] G_{21}(t_1,t_2) 
+\frac{x_1'}{2} F(t_1,t_2) &=& 0 
\nonumber \\
\Bigl[ t_1^2 \partial_1 + t_2^2\partial_2 + (x_1+\xi_1) t_1 + (x_2+\xi_2) t_2 \Bigr] G_{21}(t_1,t_2) 
+ (x_1'+\xi_1') t_1 F(t_1,t_2) &=& 0 
\label{3.2} \\[0.20cm]
\left[ t_1 \partial_1 + t_2\partial_2 +\demi(x_1+x_2) \right] H(t_1,t_2)  
+\frac{x_1'}{2} G_{12}(t_1,t_2) +\frac{x_2'}{2} G_{21}(t_1,t_2) &=& 0 
\nonumber \\
\Bigl[ t_1^2 \partial_1 + t_2^2\partial_2 + (x_1+\xi_1) t_1 + (x_2+\xi_2) t_2 \Bigr] H(t_1,t_2) & & 
\nonumber \\
+ (x_1'+\xi_1') t_1 G_{12}(t_1,t_2)+ (x_2'+\xi_2') t_2 G_{21}(t_1,t_2) &=& 0 
\nonumber
\EEA
where $\partial_i=\partial/\partial t_i$. 
One expects an unique solution, up to normalisations. It is convenient to
solve the system (\ref{3.2}) via the ansatz, with $y:=t_1/t_2$ 
\BEA
F(t_1,t_2)      &=& t_2^{-(x_1+x_2)/2}\: y^{\xi_2 +(x_2-x_1)/2} (y-1)^{-(x_1+x_2)/2-\xi_1-\xi_2} f(y) 
\nonumber \\
G_{12}(t_1,t_2) &=& t_2^{-(x_1+x_2)/2}\: y^{\xi_2 +(x_2-x_1)/2} (y-1)^{-(x_1+x_2)/2-\xi_1-\xi_2} 
\sum_{j\in\mathbb{Z}} \ln^j t_2 \cdot g_{12,j}(y) \nonumber \\
G_{21}(t_1,t_2) &=& t_2^{-(x_1+x_2)/2}\: y^{\xi_2 +(x_2-x_1)/2} (y-1)^{-(x_1+x_2)/2-\xi_1-\xi_2} 
\sum_{j\in\mathbb{Z}} \ln^j t_2 \cdot g_{21,j}(y)  \label{3.3} \\
H(t_1,t_2)      &=& t_2^{-(x_1+x_2)/2}\:  y^{\xi_2 +(x_2-x_1)/2} (y-1)^{-(x_1+x_2)/2-\xi_1-\xi_2} 
\sum_{j\in\mathbb{Z}} \ln^j t_2 \cdot h_{j}(y) \nonumber 
\EEA
{\bf 1.} The function $F$ does not contain any logarithmic 
contributions and its scaling function satisfies the
equation $f'(y)=0$, hence  
\BEQ \label{3.4}
f(y)=f_0=\mbox{\rm cste.}
\EEQ
This reproduces the well-known form of non-logarithmic local scaling \cite{Henkel06a}. 

Comparing this with the usual form (\ref{R}) of standard {\sc lsi} with $z=2$, the ageing exponents
$a,a',\lambda_R$ are related to the scaling dimensions as follows:
\BEQ
a =\demi(x_1+x_2) -1 \;\; , \;\; a'-a = \xi_1 + \xi_2 \;\; , \;\; \lambda_R = 2 (x_1 +\xi_1)
\EEQ
For example, the exactly solvable $1D$ kinetic Ising model with 
Glauber dynamics at zero temperature \cite{Godreche00a} 
satisfies (\ref{R}) with the values $a=0, a'-a=-\demi, \lambda_R=1, z=2$ \cite{Picone04}. 
Further examples of systems well described by {\sc lsi} with $a'-a\ne 0$ are given by the 
non-equilibrium critical dynamics of the kinetic Ising model with Glauber dynamics, 
both for $d=2$ and $d=3$ \cite{Henkel06a,Henkel10}; 
or the critical three-states voter-Potts model \cite{Chatelain11}.  

\noindent
{\bf 2.} Next, we turn to the function $G_{12}$. Co-variance under $X_0$ leads to the condition
\BEQ
\left( g_{12,1}(y)+\demi x_2' f(y)\right) + \sum_{j\ne 0} (j+1)\ln^j t_2 \cdot g_{12,j+1}(y) = 0
\EEQ
which must hold true for all times $t_2$. This implies
\BEQ
g_{12,1}(y) = - \demi x_2' f(y) \;\; , \;\; g_{12,j}(y) = 0 \mbox{\rm ~~;~ $\forall j\ne 0,1$}
\EEQ
In order to simplify the notation for later use, we set
\BEQ \label{3.9}
g_{12}(y) := g_{12,0}(y) \;\; , \;\; \gamma_{12}(y) := g_{12,1}(y) = -\demi x_2' f(y)
\EEQ
and these two give the only non-vanishing contributions in the ansatz (\ref{3.3}). 
Furthermore, the last remaining function $g_{12}$ is found from 
the co-variance under $X_1$, which gives
\BEQ
\sum_{j\in\mathbb{Z}} \ln^j t_2 \Bigl( y(y-1) g_{12,j}'(y) 
+ (j+1) g_{12,j+1}(y) \Bigr) + (x_2'+\xi_2') f(y) = 0 
\EEQ
for all times $t_2$. Combining the resulting two equations for 
$g_{12}$ and $\gamma_{12}$ with (\ref{3.9}) leads to
\BEQ \label{3.11}
y(y-1) g_{12}'(y) 
+ \left(\frac{x_2'}{2} +\xi_2'\right) f(y) = 0
\EEQ
{\bf 3.} The function $G_{21}$ is treated similarly. We find
\BEQ \label{3.12}
g_{21}(y) := g_{21,0}(y) \;\; , \;\; \gamma_{21}(y) := g_{21,1}(y) = -\demi x_1' f(y) \;\; , \;\;
g_{21,j}(y)=0 \mbox{\rm ~~;~ for all $j\ne 0,1$}
\EEQ
and the differential equation 
\BEQ \label{3.13}
y(y-1) g_{21}'(y) 
+ \left(x_1' +\xi_1'\right) yf(y) -\demi x_1' f(y) = 0
\EEQ
{\bf 4.} Finally, dilatation-covariance of the function $H$ leads to 
$h_j(y)=0$ for all $j\ne 0,1,2$ and
\BEA
h_1(y) &=& -\demi \bigl( x_1' g_{12}(y) + x_2' g_{21}(y) \bigr) \nonumber \\
h_2(y) &=& \frac{1}{4} x_1' x_2' f(y) \label{3.14}
\EEA
The last remaining function $h_0(y)$ is found from co-variance under $X_1$ which leads to
\BEQ
y(y-1) h_0'(y) 
+ \left(\left(x_1' +\xi_1'\right)y - \demi x_1'\right)g_{12}(y) 
+ \left( \demi x_2' + \xi_2'\right) g_{21}(y) = 0 \label{3.15}
\EEQ

Using (\ref{3.4}), the equations (\ref{3.11},\ref{3.13},\ref{3.15}) are readily solved, hence 
\BEA
g_{12}(y) &=& g_{12,0} +\left(\frac{x_2'}{2}+\xi_2'\right) f_0 \ln \left|\frac{y}{y-1}\right| 
\nonumber \\
g_{21}(y) &=& g_{21,0} -\left(\frac{x_1'}{2}+\xi_1'\right) f_0 \ln |y-1| - \frac{x_1'}{2} f_0 \ln |y| 
\nonumber \\
h_0(y) &=& h_0 - \left[ \left(\frac{x_1'}{2}+\xi_1'\right)g_{21,0} + 
\left(\frac{x_2'}{2}+\xi_2'\right)g_{12,0}\right]\ln|y-1| 
- \left[ \frac{x_1'}{2} g_{21,0} - \left(\frac{x_2'}{2}+\xi_2'\right)g_{12,0}\right]\ln|y| 
\nonumber \\
& &  + \demi f_0 \left[ \left( \left(\frac{x_1'}{2} +\xi_1'\right)\ln |y-1| 
+ \frac{x_1'}{2}\ln |y|\right)^2 
-  \left(\frac{x_2'}{2} +\xi_2'\right)^2 \ln^2\left|\frac{y}{y-1}\right| \right]  \label{3.16}
\EEA
where $f_0, g_{12,0}, g_{21,0}, h_0$ are normalisation constants. Summarising:
\BEA
F(t_1,t_2) &=& t_2^{-(x_1+x_2)/2}\: y^{\xi_2 +(x_2-x_1)/2} (y-1)^{-(x_1+x_2)/2-\xi_1-\xi_2} f_0 
\nonumber \\
G_{12}(t_1,t_2) &=& t_2^{-(x_1+x_2)/2}\: y^{\xi_2 +(x_2-x_1)/2} (y-1)^{-(x_1+x_2)/2-\xi_1-\xi_2} 
\Bigl( g_{12}(y) + \ln t_2 \cdot \gamma_{12}(y) \Bigr) 
\nonumber \\
G_{21}(t_1,t_2) &=& t_2^{-(x_1+x_2)/2}\: y^{\xi_2 +(x_2-x_1)/2} (y-1)^{-(x_1+x_2)/2-\xi_1-\xi_2} 
\Bigl( g_{21}(y) + \ln t_2 \cdot \gamma_{21}(y) \Bigr)   
\nonumber \\
H(t_1,t_2) &=& t_2^{-(x_1+x_2)/2} \: y^{\xi_2 +(x_2-x_1)/2} (y-1)^{-(x_1+x_2)/2-\xi_1-\xi_2} 
\label{3.17} \\
& & \times \Bigl( h_0(y) + \ln t_2 \cdot h_1(y) +\ln^2 t_2 \cdot h_2(y) \Bigr) \nonumber
\EEA
where the scaling functions, depending only on $y=t_1/t_2$, are given by 
eqs.~(\ref{3.9},\ref{3.12},\ref{3.14},\ref{3.16}). \\

Although the algebra $\mathfrak{age}(d)$ was written down for a dynamic exponent $z=2$, 
the space-independent part of the
two-point functions is essentially independent of this feature. The change 
$(x,x',\xi,\xi')\mapsto \bigl((2/z) x, (2/z) x', (2/z) \xi, (2/z)\xi'\bigr)$ in eq.~(\ref{3.17}) and 
eqs.~(\ref{3.9},\ref{3.12},\ref{3.14},\ref{3.16}), for both scaling operators,  
produces the form valid for an arbitrary dynamical exponent $z$. 
This observation will be used in the next section when discussing some applications. 

Since for $z=2$, the space-dependent part of the generators is not affected by the passage to the 
logarithmic theory via the
substitution (\ref{2.2}), one recovers the same space-dependence 
as for the non-logarithmic theory with $z=2$. 
For example, 
\BEA
F(t_1,t_2;\vec{r}_1,\vec{r}_2) &=& \delta({\cal M}_1+{\cal M}_2)\,\Theta(t_1-t_2) 
\, t_2^{-(x_1+x_2)/2} f_0 \nonumber \\
& & \times y^{\xi_2 +(x_2-x_1)/2} (y-1)^{-(x_1+x_2)/2-\xi_1-\xi_2}  
\exp\left[ -\frac{{\cal M}_1}{2} \frac{(\vec{r}_1-\vec{r}_2)^2}{t_1-t_2}\right] \label{3.18}
\EEA
where we also included the causality condition $t_1>t_2$, 
expressed by the Heaviside function $\Theta$, which 
can be derived using the methods of \cite{Henkel03a,Henkel12a}. 
Similar forms hold true for $G_{12}, G_{21}, H$. 

Comparison with the result (\ref{1.8},\ref{1.9}) 
of logarithmic Schr\"odinger- or conformal galilean-invariance shows:
\begin{enumerate}
\item Logarithmic contributions may arise, either as corrections 
to the scaling behaviour via additional powers of $\ln t_2$, or else
through logarithmic terms in the scaling functions themselves. 
These can be described independently in terms of the parameter sets $(x_1',x_2')$ and
$(\xi_1',\xi_2')$. 

In particular, it is possible to have representations of 
$\mathfrak{age}(d)$ with an explicit doublet in only one of the
two generators $X_0$ and $X_1$. 
\item Logarithmic corrections to scaling arise if either $x_1'\ne0$ or $x_2'\ne 0$, 
but the absence of time-translation-invariance
allows for the presence of quadratic terms in $\ln t_2$. 
\item If one sets $x_1'=x_2'=0$, 
there is no breaking of dynamical scaling through logarithmic corrections. 
However, the scaling functions 
$g_{12}(y), g_{21}(y)$ and $h_0(y)$ may still contain logarithmic terms. 

This is qualitatively distinct from logarithmic Schr\"odinger-invariance (\ref{1.8}): for example 
$H(t_1,t_2;\vec{0}) = 
\delta_{x_1,x_2}\,t_2^{-x_1} \left( H_0 - G_0 \ln (y-1) - G_0 \ln t_2 \right) (y-1)^{-x_1}$, 
with $y=t_1/t_2>1$. 
In  that case, logarithmic corrections to scaling, parametrised by $G_0$, 
are coupled to a corresponding term in the scaling function itself. 
Evidently, an analogous result holds for the logarithmic $\mbox{\sc cga}$. 
\item The constraint $F=0$ of both logarithmic conformal invariance and logarithmic
Schr\"o\-din\-ger/con\-for\-mal galilean invariance is no longer required. 
\item If time-translation-invariance is assumed, one has 
$\xi_1=\xi_2=\xi_1'=\xi_2'=0$, $x_1=x_2$ and $f_0=0$. 
The functional form of eqs.~(\ref{3.17},\ref{3.18}) 
then reduces to the Schr\"odinger-invariant forms of eq.~(\ref{1.8}).  
\end{enumerate}

\section{Applications} 

Several candidate model systems for an application of logarithmic {\sc lsi} 
({\sc llsi}) in physical ageing will be discussed. The models analysed here, 
namely the universality classes of the Kardar-Parisi-Zhang equation and directed percolation, are widely considered to
be the most simple models for the non-equilibrium phase transitions they describe -- and in this
sense play about the same r\^ole as the Ising model in equilibrium critical phenomena. It has been
established in recent years that they undergo ageing in the sense that the three defining properties 
listed in the introduction hold true, see e.g.
\cite{Kall99,Dornic01,Enss04,Ramasco04,Daqu11,Henk12,Hyun12}.  

\subsection{One-dimensional Kardar-Parisi-Zhang equation}

An often-studied situation is the growth of interfaces, 
which on a lattice may be described in terms
of time-dependent heights $h_i(t)\in\mathbb{N}$ (and $i\in\mathbb{Z}$), 
and subject to a stochastic deposition of particles. 
If one further admits a RSOS constraint of the form 
$0\leq |h_{i+1}(t)-h_i(t)|\leq 1$ \cite{Kim89}, this
goes in a continuum limit to the paradigmatic model 
equation proposed by Kardar, Parisi and Zhang (KPZ) 
\cite{Kard86}, described by a
time-dependent height variable $h=h(t,{r})$
\BEQ \label{kpz}
\frac{\partial h}{\partial t} = \nu \frac{\partial^2 h}{\partial {r}^2} 
+ \frac{\mu}{2} \left( \frac{\partial h}{\partial {r}}\right)^2 +\eta
\EEQ
where $\eta(t,{r})$ is a white noise with zero mean and variance 
$\langle\eta(t,{r})\eta(t',{r}')\rangle=2\nu T\delta(t-t') \delta({r}-{r}')$ 
and $\mu,\nu,T$ are material-dependent constants. 
Its many applications include Burgers turbulence, 
directed polymers in a random medium, glasses and
vortex lines, domain walls and biophysics, 
see e.g. \cite{Bara95,Halp95,Krug97,Krie10,Sasa10b,Tong94,Batc00,Corwin11}
for reviews. In $1D$ the height distribution can be shown to converge for large times 
towards the gaussian Tracy-Widom distribution \cite{Sasa10,Cala11,Geudre12}. 
Experiments on the growing interfaces of turbulent liquid crystals 
reproduce this universality class \cite{Take11}. 

Physical ageing of two-time quantities in this universality class having been 
studied several times in the past \cite{Kall99,Bust07,Chou10,Daqu11,Krec97,Henk12}; 
here we concentrate exclusively on the linear response of the height $h_i(t)$ with respect to the
local particle-deposition rate $p_i(t)$, viz. 
$R(t,s) = \left.\delta \langle h_i(t)\rangle/\delta p_i(s)\right|_{p=0}$. 
In practise, an integrated response can be defined for the discrete-height model \cite{Kim89} 
by considering a space-dependent
deposition rate $p_i=p_0 + a_i \eps /2$ with $a_i=\pm 1$ and $\eps=0.005$ a small parameter. 
Then consider, with the {\em same} stochastic noise $\eta$, two
realisations: system A evolves, up to the waiting time $s$, 
with the site-dependent deposition rate $p_i$ and 
afterwards, with the uniform deposition rate $p_0$. 
System B evolves always with the uniform deposition rate $p_i=p_0$.
Then, the time-integrated autoresponse function is
\BEQ \label{4.2}
\chi(t,s) = \int_0^s \!\!\D u\: R(t,u) = 
\frac{1}{L} \sum_{i=1}^L \left\langle \frac{h_{i}^{(A)}(t;s) -
h_{i}^{(B)}(t)}{\eps a_i}\right\rangle  
= s^{-a} f_{\chi}\left( \frac{t}{s}\right)
\EEQ
together with the generalised Family-Vicsek scaling \cite{Henk12}. 
The autoresponse exponent is read off from 
$f_{\chi}(y)\sim y^{-\lambda_R/z}$ for $y\to\infty$. In $1D$, one has the well-known exponents
$a=-1/3$, $\lambda_R=1$ and $z=3/2$. 

\begin{figure}[tb]
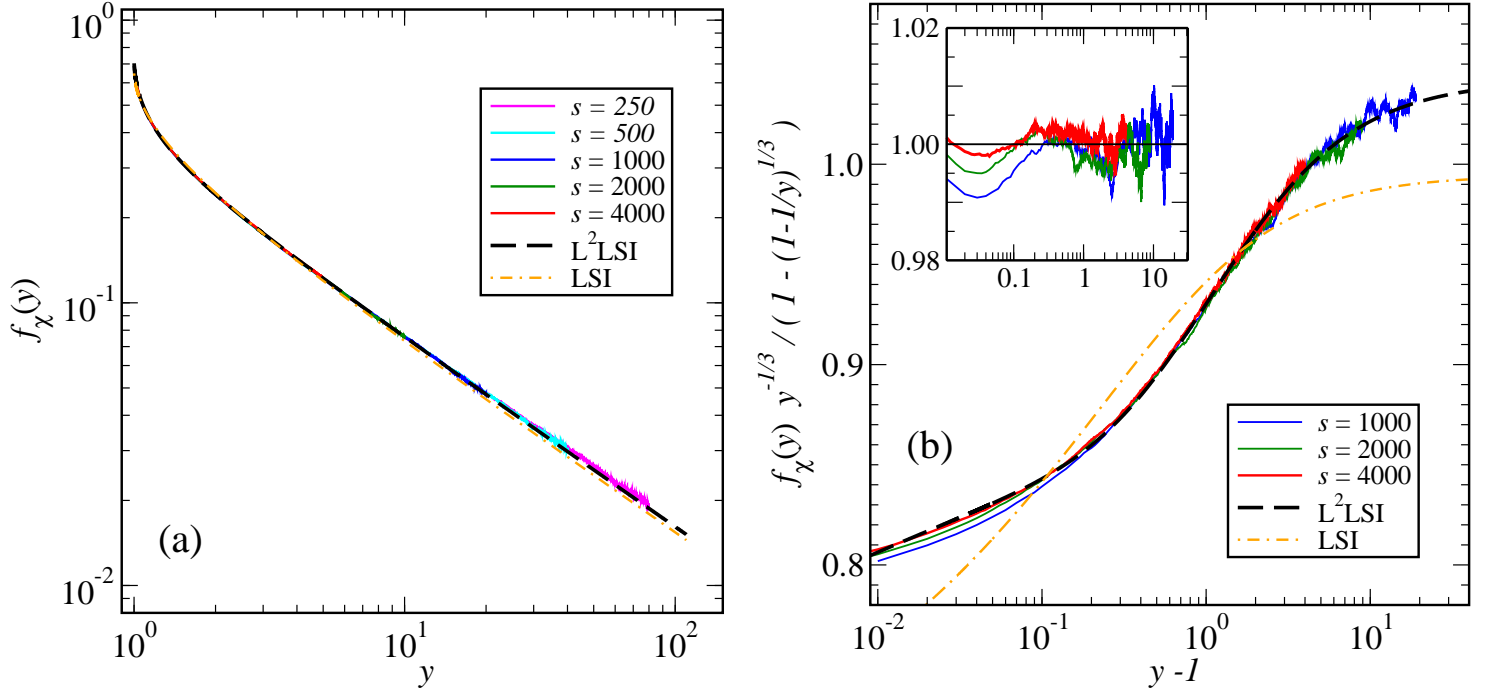

\centerline{\psfig{figure=vieuxlog_abb2a.eps,width=3.75in,clip=} ~ \psfig{figure=vieuxlog_abb2b.eps,width=3.75in,clip=}}
\caption{\label{fig2} Scaling of the integrated autoresponse $\chi(t,s) = s^{+1/3} f_{\chi}(t/s)$ of the $1D$
Kardar-Parisi-Zhang equation, as a function of $y=t/s$, for several values of the waiting time $s$. 
(a) Standard scaling plot of $f_{\chi}(y)$ over against $y$.  
(b) Scaling of the reduced scaling function 
$f_{\rm red}(y) = f_{\chi}(y) y^{-1/3} \left[1-(1-y^{-1})^{1/3}\right]^{-1}$.  
The dash-dotted line labelled {\sc lsi} gives a fit to non-logarithmic {\sc lsi} (see text) and the 
dashed line labelled {\sc l$^2$lsi} gives the prediction (\ref{4.5},\ref{4.6}). 
The inset in (b) displays the ratio $f_{\chi}(y)/f_{\rm L^2LSI}(y)$ over against $y$. 
The data are from \cite{Henk12}.}
\end{figure}

In figure~\ref{fig2}a, the resulting scaling behaviour (\ref{4.2}) of data for the autoresponse 
as obtained from intensive numerical simulations \cite{Henk12} is shown (generated from an initial flat surface). 
There is a clear data collapse for sufficiently large values of $s$ and the data clearly confirm the expected values
of the ageing exponent $a=-\frac{1}{3}$ and, from the power-law decay for $y\gg 1$, the autoresponse exponent
$\lambda_R/z=\frac{2}{3}$ \cite{Kall99,Bust07,Daqu11,Krec97,Henk12}. The data can be compared successfully with the
prediction (\ref{R}) of non-logarithmic {\sc lsi}, with the estimated exponent $a'\simeq -0.5$. Although in this kind 
of plot the agreement between the data and {\sc lsi} appears to work very well, it has been realised in recent years that
there are better and more meaningful ways to test the agreement of numerical data with theoretical shapes, such as predicted
by {\sc lsi}, in a much more precise way. In this way, it has turned out that when data for increasingly larger values of
$s$ can be obtained, increasingly finer details in the shape of the scaling function for values 
$y\approx 1$ must be taken into account. A first step in our slowly improving understanding of the shapes of these scaling functions
had been the observation that $a'-a\ne 0$ in general (which distinguishes the predictions of $\mathfrak{age}(d)$-invariance from
those of $\mathfrak{sch}(d)$-invariance) \cite{Picone04,Henkel06a,Odor06,Lorenz07,Chatelain11}. As we shall show below, it
turns out that plotting data as in figure~\ref{fig2}a is not yet sufficient to reliably analyse finer details of the
shape of $f_{\chi}(y)$ in the limit $y\to 1^{+}$.

We propose to use {\sc llsi} for that purpose. 
In order to test eq.~(\ref{3.17}) (with the tacit extension to generic $z$ as outlined above) in the
$1D$ KPZ universality class, we make the working hypothesis
$R(t,s) = \langle \psi(t)\wit{\psi}(s)\rangle$, where the two scaling operators $\psi$ and $\wit{\psi}$ 
are described by the logarithmically extended scaling dimensions
\BEQ
\left(\matz{x}{x'}{0}{x}\right) \;\; , \;\; 
\left(\matz{\xi}{\xi'}{0}{\xi}\right) \;\; \mbox{\rm ~~and~~} \;\; 
\left(\matz{\wit{x}}{\wit{x}'}{0}{\wit{x}}\right) \;\; , \;\; 
\left(\matz{\wit{\xi}}{\wit{\xi}'}{0}{\wit{\xi}}\right) 
\EEQ 
In principle, one might have logarithmic corrections to scaling, according to eq.~(\ref{3.17}).
However, we interpret the clear data collapse in figure~\ref{fig2}ab 
as evidence that no such corrections should arise.
Hence the two functions $h_{1,2}(y)$ must vanish. 
Because of eq.~(\ref{3.14}), this means that $x'=\wit{x}'=0$. Furthermore, the requirement of a simple
power-law for $y\gg 1$, implies $\xi'=0$ from the explicit form (\ref{3.16}) of $h_0(y)$. 
Logarithmic representations of {\sc lsi} are then described by
$\wit{\xi}'$  only, which can always be normalised to $\wit{\xi}'=1$. 
If we take 
$R(t,s) = \left\langle \psi(t)\wit{\psi}(s)\right\rangle=s^{-1-a} f_R(t/s)$, it remains 
\BEQ \label{4.4}
f_R(y) = y^{-\lambda_R/z} \left( 1 - y^{-1}\right)^{-1-a'} 
\left[ h_0 - g_0 \ln\left( 1 - y^{-1}\right) - \demi f_0  \ln^2\left( 1 - y^{-1}\right)\right] 
\EEQ
with the exponents $1+a = (x + \wit{x})/z$, $a'-a = \frac{2}{z} \left( \xi + \wit{\xi}\,\right)$, 
$\lambda_R/z = x + \xi$ and the normalisation constants $h_0, g_0=g_{12,0},f_0$. 
Using the specific value $\lambda_R/z-a=1$ which holds for the $1D$ KPZ, the integrated autoresponse 
$\chi(t,s) = s^{-a} f_{\chi}(t/s)$ becomes  
\BEQ \label{4.5} 
f_{\chi}(y) = y^{+1/3} 
\left\{ A_0 \left[ 1 - \left( 1- y^{-1}\right)^{-a'} \right] \right.  
+ \left.  \left( 1 - y^{-1}\right)^{-a'} \left[ A_1 \ln\left( 1 - y^{-1}\right) 
+ A_2  \ln^2\left( 1 - y^{-1}\right) \right] \right\}
\EEQ
where $A_{0,1,2}$ are normalisations related to $f_0,g_0,h_0$. Indeed, for $y\gg 1$, 
one has $f_{\chi}(y) \sim y^{-2/3}$, as expected. 
The non-logarithmic case would be recovered for $A_1=A_2=0$. 

In figure~\ref{fig2}b, the simulational data from \cite{Henk12}  
are compared with the predicted form (\ref{4.5}). 
Since we are interested in finer features of the scaling function $f_{\chi}(y)$,  
and in order to be able to distinguish a non-trivial shape of 
$f_{\chi}(y)$ from the omnipresent finite-time corrections,  
very large values of the waiting time $s$ must be considered. 
This is especially the case for values $y\approx 1$, where deviations
of $f_{\chi}(y)$ from the asymptotic power-law 
are the strongest but also the finite-time corrections become maximal. 
The form chosen here for the scaling plot is selected 
for a good sensitivity to the shape of $f_{\chi}(y)$. 

Although we have already observed a good data collapse, we also observe from figure~\ref{fig2}b 
that data with $s<10^3$ are not yet fully in the scaling regime. 
Still, we conclude that logarithmic corrections to scaling should be unimportant.  
The chosen plot readily permits several tests. 
First, if non-logarithmic {\sc lsi} with the extra hypothesis $a=a'$ would hold, 
one should observe $f_{\rm red}(y)=\mbox{\rm cste.}$,
which clearly is not the case. Second, a much better agreement is found if $a'$ 
is allowed to differ from $a$. The dashed-dotted curve labelled `{\sc lsi}'
in figure~\ref{fig2}b, with an assumed value $a'=-0.5$, shows that while 
the data can be described by non-logarithmic {\sc lsi} with an accuracy of about $5\%$, the earlier plot
in figure~\ref{fig2}a did not permit to detect such differences. Third, and interestingly, 
if one tries to include only the first of the logarithmic terms in the scaling function (\ref{4.5}) by
constraining $A_2=0$, the best fit cannot be distinguished from the non-logarithmic one, with an estimate 
$|A_1| \lesssim 6 \cdot 10^{-4}$. 
Finally, only if one uses the full structure of logarithmic {\sc lsi},
an excellent representation of the data is found, labelled `{\sc l$^2$lsi}' in figure~\ref{fig2}b,
and to an accuracy better than $0.1\%$ over the range of data available. 
A least-squares fit leads to the estimates \cite{Henk12}
\BEQ \label{4.6}
a'=-0.8206\;\; , \;\; A_0 = 0.7187\;\; , \;\; A_1 = 0.2424\;\;,\;\; A_2 = -0.09087
\EEQ
This fit should be meaningful since all amplitudes are of a comparable order of magnitude. 
In the inset the ratio $\chi(t,s)/\chi_{\mbox{\rm\footnotesize L$^2$LSI}}(t,s)$ 
is shown and we see that at least down
to $t/s\approx 1.03$, the data collapse indicating dynamical scaling holds true, 
within  the accuracy limits set by the
stochastic noise, within $\approx 0.5\%$. 
For the largest waiting time $s=4000$, this observation extends over the
entire range of values of $t/s$ considered.

\subsection{One-dimensional critical directed percolation} 

The directed percolation universality class is the paradigmatic example of a non-equilibrium phase transition
with an absorbing state. It has been realised in countless different ways, 
with often-used examples being either the contact process or else Reggeon field theory, 
and very precise estimates of the location of the critical point and the critical exponents are known, 
see \cite{Hinrichsen00,Odor04,Henkel09} and references therein. 
Its predictions are also in agreement with extensive recent experiments in turbulent liquid crystals \cite{Takeuchi09}.  
Since it is well-understood that critical $2D$ {\em isotropic}
percolation can be described in terms of conformal invariance 
\cite{Langlands94},\footnote{Cardy \cite{Cardy92} 
and Watts \cite{Watts96} used conformal invariance to derive their celebrate formul{\ae} 
for the crossing probabilities. A precise formulation of the conformal invariance methods 
required in their derivations actually leads to a 
logarithmic conformal field theory \cite{Mathieu07}.} one might wonder whether
some kind of local scale-invariance might be applied to directed percolation.

\begin{figure}[tb]
\centerline{\psfig{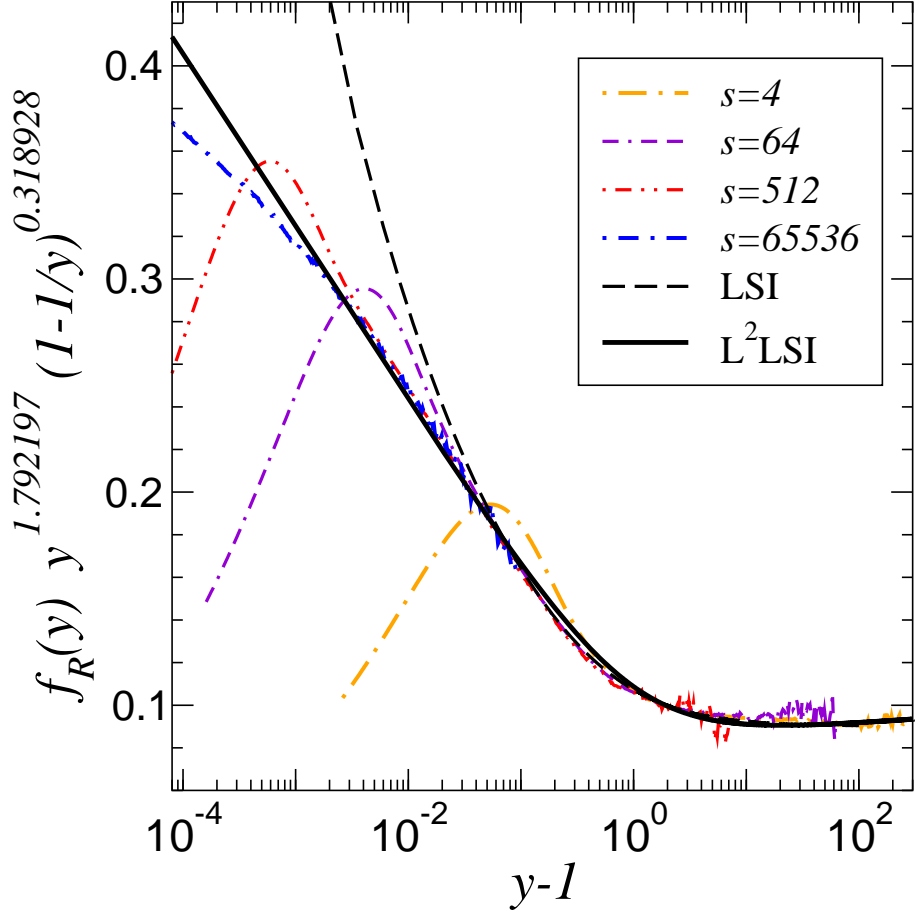}}
\caption{\label{fig1} Reduced scaling function $h_R(y) = f_R(y) y^{\lambda_R/z} (1-1/y)^{1+a}$ 
the autoresponse $R(t,s)=s^{-1-a}f_R(t/s)$ 
of the $1D$ critical contact process, as a function of $y=t/s$, for several
values  of the waiting time $s$.  
The dashed line labelled `{\sc lsi}' is from (\ref{R}), with $a'-a=0.26$. The full curve labelled
`{\sc l$^2$lsi}' is obtained from eq.~(\ref{4.8}), derived from logarithmic {\sc lsi} with $f_0=0$,
and the parameters (\ref{4.9}), see text.}
\end{figure}

In the contact process, a response function can be defined by considering the 
response of the time-dependent particle concentration with respect to a time-dependent 
particle-production rate. The relaxation from an initial state is in many respects 
quite analogous to what is seen in systems with an equilibrium stationary state 
\cite{Enss04,Ramasco04,Baumann07}. 
In figure~\ref{fig1}, we show simulational data of the autoresponse function
$R(t,s)=s^{-1-a} f_R(t/s)$ of $1D$ critical directed percolation, 
realised here by the critical contact process and 
as initial state uncorrelated particles at a finite density \cite{Enss04,Enss}. 
Plotting directly the scaling function $f_R(y)$ over against $y=t/s$ has led to a very good agreement of the
data with non-logarithmic {\sc lsi}, with $a'-a\simeq 0.27$ \cite{Enss04,Ramasco04}, quite analogously to
figure~\ref{fig2}a above. 
In order to study the shape of the scaling function in detail, especially for $y\to 1^{+}$, consider
\BEQ
h_R(y) := f_R(y) y^{\lambda_R/z} (1-y^{-1})^{1+a}
\EEQ
with the exponents taken from \cite{Henkel09}. 
We see in figure~\ref{fig1} that while for $y=t/s$ large enough, the data collapse is excellent, 
finite-time corrections become increasingly more
important when $y$ is lowered toward unity. 

The definition of $h_R(y)$ permits several tests of {\sc lsi} on different levels of precision, beginning at large values of
$y$ and proceedings towards $y\to 1$. 
First, a non-logarithmic form with the extra assumption $a=a'$ would from (\ref{R}) 
lead to a constant from $h_R(y)=\mbox{\rm cste.}$, which only
describes the data for $y\gtrsim 3-4$. 
Second, a better fit, which assumes $a'-a=0.26$ describes the data down to 
$y\approx 1.1$, is obtained when $a'$ is allowed to be fitted to the
data \cite{Henkel06a}. Still, further systematic deviations exist when 
$t/s$ is yet closer to unity and we shall now try to
use logarithmic {\sc lsi} in order to account for the data. 

Again, we propose to use {\sc llsi}. We make the working assumption 
$R(t,s)=\left\langle \psi(t)\wit{\psi}(s)\right\rangle$ and interpret the good quality of the
data collapse as evidence for the absence of logarithmic corrections to scaling. 
This implies that $x'=\wit{x}'=0$.  
Then logarithmic {\sc lsi} eq.~(\ref{3.16}) predicts
\BEA
h_R(y) &=&  \left( 1 - \frac{1}{y}\right)^{a-a'} 
\left( h_0 - g_{12,0} \wit{\xi}'  \ln (1-1/y) - \demi f_0 \wit{\xi}'^2 \ln^2 (1-1/y)
\right. \nonumber \\
& & \left. ~~~~~~- g_{21,0} \xi'  \ln (y-1) + \demi f_0 \xi'^2 \ln^2 (y-1) \right) 
\label{4.8}
\EEA
Further constraints must be obeyed, in particular 
the resulting scaling function should always be positive. 

Numerical experiments reveal that the best fits are 
obtained by fitting the generic form (\ref{4.2}) to the data. It then
turns out that the terms which depend quadratically 
on the logarithms have amplitudes which are about $10^{-4}$ times smaller
than those of the other terms. We consider this as evidence that $f_0=0$. 
Making this assumption, one has the 
phenomenological scaling form 
$h_R(y)= h_0 (1-1/y)^{a-a'} \left( 1 - (A+B)\ln(1-1/y) +B\ln (y-1)\right)$,
where $h_0$ is a normalisation constant and 
$A,B$ are two positive universal parameters. The best fit is found if 
\BEQ \label{4.9}
a-a'=0.00198 \;\; , \;\; A=0.407 \;\; , \;\; B=0.02 \;\; , \;\; h_0 = 0.08379
\EEQ
and gives a good description of the data, down to 
$y-1\approx 2 \cdot 10^{-3}$ (for smaller values of $y$, we cannot be sure to be still
in the scaling regime). 

Note that our current estimate $a'-a\simeq -0.002$ 
is quite distinct from the earlier estimate $a'=a\approx 0.27$ \cite{Henkel06a}
and also implies a small logarithmic contribution in the $y\gg 1$ limit.

\section{Conclusions}

We have discussed the extension of dynamical scaling 
towards local scale-invariance in the case when the
physical scaling operator acquires a single `logarithmic' partner with the same scaling dimension. 
Since in far-from-equilibrium relaxation, time-translation-invariance
does not hold, one cannot appeal directly to the 
known cases of logarithmic conformal, 
logarithmic Schr\"odinger- or logarithmic conformal galilean invariance. 
Indeed, analogously to the non-logarithmic case, the doublets of scaling operators 
are described by {\em pairs} of Jordan matrices of the {\em two} distinct 
and independent scaling dimensions of each quasi-primary scaling operator. 
When computing two-point functions transforming co-variantly under logarithmic representations of the
algebra $\mathfrak{age}(d)$, the absence of 
time-translation-invariance renders independent logarithmic corrections to scaling and 
also non-trivial logarithmic modification of the scaling functions, 
see eqs.~(\ref{3.16},\ref{3.17}). These results
generalise the forms found from logarithmic Schr\"odinger-invariance \cite{Hosseiny10}. 

These predictions have been compared to simulational data in two 
non-equilibrium model systems undergoing physical ageing,
namely the $1D$ Kardar-Parisi-Zhang equation and $1D$ 
critical directed percolation. 
A close analysis of the shape of the scaling function of the 
linear autoresponse of the order parameter revealed systematic
deviations of the numerical data from the predictions of non-logarithmic {\sc lsi}, 
even if the exponent $a'\ne a$ is introduced as
a further free parameter. On the other hand, 
logarithmic {\sc lsi} fits the available data well, and over the entire range of
the scaling variable $y=t/s$ for which numerical data were available. 

However, the large number 
of undetermined normalisation constants gives a considerable flexibility to these fits. 
It remains an open question if logarithmic {\sc lsi} 
might be construed in a way which would produce more constraints 
between these so far independent normalisation constants. 
Finding an exactly solvable example of {\sc llsi} is another {\it desideratum}.  
It is conceivable that the logarithmic terms found 
in the scaling function in the simple phenomenological scheme proposed here 
are but the first few terms of an infinite logarithmic series, 
perhaps in analogy to ideas raised long ago in \cite{Gribov81,Gurarie93}. 
Of course, further independent tests of the proposal presented here would be desirable. 

In a sense, since ordinary critical $2D$ percolation is described in terms of {\em  logarithmic} conformal
invariance, such that there must exist a logarithmic partner to the physical order parameter (still unidentified to
the best of our knowledge) \cite{Mathieu07}, 
it might appear natural that a similar phenomenon should also occur for directed percolation. 
It remains an important open question how to physically identify the logarithmic partners whose effects seem to be
present in the shape of the autoresponse scaling function. Given the quite distinct nature of the two universality classes
studied in this work, it is conceivable that analogous findings may hold true in other models as well, for example
in $2D$ critical majority voter models \cite{Henkel13}. 

Since logarithmic conformal theories are thought 
to be closely related to non-local observables \cite{Cardy92,Watts96,Mathieu07}, one might also wonder
whether the empirical observation of {\sc llsi} might be indicative of some sort of non-locality. Possibly, there might
exist a link with the celebrate scaling relations which link the global persistence exponent $\theta_g$ with the
autoresponse/autocorrelation exponent of ageing and equilibrium critical exponents,\footnote{This exponent describes the
long-time decay of the probability $P_g(t)\sim t^{-\theta_g}$ that the global order parameter has not changed its sign until time $t$.}  
and which can be derived both at criticality \cite{Majumdar96a}
and in the entire ordered phase \cite{Cueille97,Henkel09a}. 
These relations depend in their derivation on the assumption that the global order parameter
is gaussian and that even after renormalisation, its long-time dynamics is {\em markovian}. 
However, these scaling relations are known
to be invalid in most systems, with the only exception of some integrable models, based on free fields 
(see \cite[ch. 1.6 \& 3.2.4]{Henkel10} for a compilation of explicit model results). 
Since in turn these scaling relations for $\theta_g$ are equivalent to a certain global correlator having a pure power-law
form, it is possible that the derivation and test of a prediction of {\sc llsi} of this correlator could illustrate this
question from a new angle. We hope to return to this question in the future.

Since logarithmic conformal invariance also arises in disordered systems at equilibrium, 
it would be of interest to see whether 
logarithmic local scale-invariance could help in improving 
the understanding of the relaxation processes of disordered systems 
far from equilibrium, see e.g. \cite{Paul04,Henkel08,Loureiro10,Park10}.

\noindent 
{\bf Acknowledgement:} I thank T. Enss for the {\sc tmrg} data, 
M. Pleimling and J.D. Noh for useful correspondence 
and the Departamento de F\'{\i}sica da Universidade de Aveiro (Portugal) for warm hospitality. 

\appsection{A}{Two-point functions in logarithmic conformal invariance}

We briefly recall the derivation of the form (\ref{1.4}) of 
the two-point functions (\ref{1.3}) -- which transform 
co-variantly under the logarithmic representations
of conformal invariance \cite{Gurarie93,Rahimi97}. 
We shall restrict to the most simple case when
a quasi-primary scaling operator 
$\Psi=\left(\vekz{\psi}{\phi}\right)$ is a doublet and also concentrate
on the left-moving part described by the variable $w$. The conformal
generators act as follows on the components
\BEA
\ell_n \phi(w) &=& \left( - w^{n+1} \partial_w - (n+1)\Delta w^n \right) \phi(w) 
\nonumber \\
\ell_n \psi(w) &=& \left( - w^{n+1} \partial_w - (n+1)\Delta w^n \right) \psi(w) - (n+1) w^n \phi(w)
\EEA 

Using the definition (\ref{1.3}) of the two-point functions, it is obvious from translation-invariance
(with generator $\ell_{-1}$) that $F=F(w)$, $G=G(w)$, $H=H(w)$ with $w=w_1-w_2$. Furthermore,
standard dynamical scaling gives $F(w) = F_0 w^{-2\Delta}$. Next, co-variance of the
`mixed' two-point function gives
$\ell_n^{[2]} G = \left\langle \left(\ell_n \phi(w_1)\right) \psi(w_2)\right\rangle 
+ \left\langle \phi(w_1) \left( \ell_n \psi(w_2)\right) \right\rangle \stackrel{!}{=} 0$, 
which gives, for $n=0,1$, respectively
\BEQ
\left( - w \partial_w - 2\Delta\right) G(w) - F(w) = 0 \;\; , \;\;
\left( - w^2 \partial_w - 2\Delta w \right) G(w) = 0
\EEQ
Combination of these yields $w F(w)=0$, hence
\BEQ
F(w) = 0 \;\; , \;\; G(w) = G_0 w^{-2\Delta}
\EEQ
Similarly, for the last two-point function one has for $n=0,1$, respectively
\BEA
\hspace{-0.8truecm}\left( - w\partial_w -2\Delta \right) H(w) - G(w) - G(-w) &=& 0 \nonumber \\  
\hspace{-0.8truecm}\left( -w^2\partial_w - 2\Delta w\right) H(w) - 2wG(w) 
+ 2w_2 \underbrace{[ ( -w\partial_w -2\Delta) H(w) - G(w) - G(-w) ]}_{=0} &=& 0 
\label{A.4}
\EEA
and where the first of these is to be used again. Combination of the two equations (\ref{A.4}) gives
$2G(w)=G(w) + G(-w)$, such that the `mixed' two-point function $G(w)$ is even
\BEQ
G(w) = G(-w) = G_0 |w|^{-2\Delta}
\EEQ
as stated in (\ref{1.4}). Integration of the remaining equation 
$\left(-w\partial_w -2\Delta\right) H(w) -2G_0 |w|^{-2\Delta}=0$ completes the derivation,
where the normalisation constants $G_0, H_0$ remain undetermined. 

The same result can be found from the formalism of nilpotent variables \cite{Moghimi00,Hosseiny10}. 

\appsection{B}{On logarithmic scaling forms} 

In the ageing of several magnetic systems, such as the 
$2D$ XY model quenched from a fully disordered initial state
to a temperature $T<T_{\rm KT}$ below the 
Kosterlitz-Thouless transition temperature \cite{Bray00,Berthier01,Abriet04a} 
or fully frustrated spin systems
quenched onto their critical point \cite{Walter08,Karsai09}, 
the following phenomenological scaling behaviour
\BEQ \label{B1}
R(t,s) = s^{-1-a} f_R\left( \frac{t}{\ln t} \frac{\ln s}{s}\right)
\EEQ
has been found to describe the simulational data well. 
Is this scaling form consistent with {\sc llsi}~?  
{\it H\'elas}, this question has to be answered in the negative.
If one fixes $y=t/s$ and expands the quotient 
$\ln s/\ln t = \ln s/(\ln y + \ln s)$ for $s\to\infty$, 
eq.~(\ref{B1}) leads to the generic scaling behaviour
\BEQ \label{s}
R(t,s) = s^{-1-a} \sum_{k,\ell} f_{k,\ell} \: y^k \left( \frac{\ln y}{\ln s}\right)^{\ell}
\EEQ
Comparison with the explicit scaling forms derived in section~3 
shows that there arise only combinations of the
form $\ln^n y \cdot \ln^m s$ or $\ln^n (y-1) \cdot \ln^m s$, 
where the integers $n,m$ must satisfy $0\leq n+m\leq 2$. This is incompatible with (\ref{s}). 

In conclusion, the logarithmic scaling form (\ref{B1}) 
cannot be understood in terms of logarithmic local scale-invariance, as presently formulated.


\end{document}